\documentclass[11pt,letterpaper]{article}
\pdfoutput=1
\usepackage{jinstpub}
\usepackage{color}
\usepackage[table]{xcolor}
\usepackage{graphicx}

\usepackage{verbatim}
\usepackage{amsmath}
\usepackage{amssymb}
\usepackage{subfig}
\usepackage{url}
\usepackage{bbold}
\usepackage{slashed}
\usepackage{url}
\usepackage{multirow}
\usepackage{threeparttable}
\usepackage{paralist}
\usepackage{bm}
\usepackage{hyperref}
\usepackage{lineno}
\hypersetup{
    colorlinks=true,
    linkcolor=cyan,
    filecolor=magenta,      
    urlcolor=blue,
}
\usepackage{siunitx}

\newcommand{\be}{\begin{equation}}
\newcommand{\ee}{\end{equation}}

\newcommand{\hlsfml}{{\tt{hls4ml}}}

\newcommand{\initialsspace}{0.1em}
 
\makeatletter
\newcommand{\splitlist}[1]{\@splitlist#1\@nil}
\def\@splitlist#1\@nil{%
  \if\relax\detokenize{#1}\relax
    \expandafter\@gobble
  \else
    \expandafter\@firstofone
  \fi
  {\@spl@tlist#1.\@nil}%
}
 
\def\@spl@tlist#1.#2\@nil{%
    \def\tmpA{#1}%
    \def\tmpB{#2}%
    \def\tmpP{.}%
    \ifx\tmpB\tmpP%
        #1.%
    \else{%
        \ifx\tmpA\@empty%  
        \else%
                #1.\nobreak\hspace{\initialsspace}%
        \fi%
    }% 
    \fi%
  \if\relax\detokenize{#2}\relax
    \expandafter\@firstoftwo
  \else
    \expandafter\@secondoftwo
  \fi
  {\unskip}%
  {\@spl@tlist#2\@nil}%
}
\makeatother
 
\newcommand{\initials}[1]{\splitlist{#1}}

\begin{document}

%\title{Deep Neural Networks in FPGAs for Trigger and DAQ}
%\title{Deep neural networks in FPGAs for real-time particle physics applications}
\title{Fast inference of Boosted Decision Trees in FPGAs for particle physics}

\author[a]{Sioni Summers}
\author[b]{, Giuseppe Di Guglielmo}
\author[c]{, Javier Duarte}
\author[d]{, Philip Harris}
\author[e]{, Duc Hoang}
\author[f]{, Sergo Jindariani}
\author[g]{, Edward Kreinar}
%\author[c]{, Mia Liu}
\author[a,h]{, Vladimir Loncar}
\author[a]{, Jennifer Ngadiuba}
%\author[c]{, Kevin Pedro}
\author[a]{, Maurizio Pierini}
\author[d]{, Dylan Rankin}
\author[f]{, Nhan Tran}
\author[i]{, Zhenbin Wu}

\affiliation[a]{European Organization for Nuclear Research (CERN), CH-1211 Geneva 23, Switzerland}
\affiliation[b]{Columbia University, New York, NY 10027, USA}
\affiliation[c]{University of California San Diego, La Jolla, CA 92093, USA}
\affiliation[d]{Massachusetts Institute of Technology, Cambridge, MA 02139, USA}
\affiliation[e]{Rhodes College, Memphis, TN 38112, USA}
\affiliation[f]{Fermi National Accelerator Laboratory, Batavia, IL 60510, USA}
\affiliation[g]{HawkEye360, Herndon, VA 20170, USA}
\affiliation[h]{Institute of Physics Belgrade, Serbia}
\affiliation[i]{University of Illinois at Chicago, Chicago, IL 60607, USA}

%\emailAdd{ntran@fnal.gov}
\emailAdd{hls4ml.help@gmail.com}

\abstract{
We describe the implementation of Boosted Decision Trees in the \texttt{hls4ml} library, which allows the translation of a trained model into FPGA firmware through an automated conversion process. Thanks to its fully on-chip implementation, \texttt{hls4ml} performs inference of Boosted Decision Tree models with extremely low latency. 
With a typical latency less than $\SI{100}{ns}$, this solution is suitable for FPGA-based real-time processing, such as in the Level-1 Trigger system of a collider experiment.
These developments open up prospects for physicists to deploy BDTs in FPGAs for identifying the origin of jets, better reconstructing the energies of muons, and enabling better selection of rare signal processes.
}

\maketitle

%%%%%%%%%%%%%%%%%%%%%%%%%%%%%%%%%%%%%%%%%%%%%%%%%%%%%%%%%%%%%%%%%%%%%%%%%%%%%%%%%%%%%%%%%%%%%%%%
% I N T R O D U C T I O N
%%%%%%%%%%%%%%%%%%%%%%%%%%%%%%%%%%%%%%%%%%%%%%%%%%%%%%%%%%%%%%%%%%%%%%%%%%%%%%%%%%%%%%%%%%%%%%%%
%\clearpage
\section{Introduction}
\label{sec:introduction}
%A new software package to translate trained neural networks into FPGA firmware capable of low latency inference, called \hlsfml, was previously presented~\cite{hls4ml}.
%In this work, we present the addition of an implementation of Boosted Decision Trees (BDTs) to \hlsfml.
Starting with the work of the MiniBooNE collaboration~\cite{Roe:2004na,Yang:2005nz}, Boosted Decision Trees (BDTs) have been extremely prevalent within the field of High Energy Physics (HEP)~\cite{Radovic:2018dip}, used mainly for regression and classification tasks, both in event reconstruction and subsequent data analysis.
In the high-profile discovery of the Higgs boson, BDTs were used to increase the sensitivity of the CMS analysis in the decay channel of the Higgs to two photons~\cite{higgs_cms}, and have been used significantly in further analyses of Higgs properties.

At the Large Hadron Collider (LHC) experiments, proton collisions occur at such a frequency that the full rate of data cannot be stored.
With the LHC delivering collisions every $\SI{25}{ns}$, the experiments CMS and ATLAS have to deal with tens of terabytes of data produced each second. Each experiment operates an online data reduction system, called the trigger, to filter out only a fraction of events for further analysis.
Due to the extreme data rates, this processing must necessarily be extremely fast, and since the rejected events can never be recovered, the selection must be highly robust. 

The CMS and ATLAS experiments deploy a two-stage trigger system, starting with the Level-1 Trigger (L1T) performing a first selection, with a second High Level Trigger (HLT) performing a more refined selection.
The L1T must process each LHC event, at the full $\SI{40}{MHz}$ collision rate, and return its decision within approximately $\SI{10}{\mu s}$, the latency for which the event data can be buffered. Due to these constraints, the L1T is implemented using high speed electronics, consisting of ASICs and FPGAs on custom cards, with high-speed optical interconnects.

Recently, Deep Neural Networks (DNNs) have been investigated as an alternative to BDTs for HEP applications~\footnote{For an extensive discussion of use cases, see Ref.~\cite{Guest:2018yhq} and references therein.}, due to their superior performance and the increasing availability of parallel processors capable of high throughput training and inference. Despite the large amount of studies showing interesting use cases for DNN applications, the number of DNN models deployed in the central data processing of the LHC experiments during previous LHC running was very limited. This was mainly due to the lack of optimal deployment solutions that would meet the strong constraints of central processing systems (e.g., real-time event selection in the trigger systems), both in terms of latency and computing resource footprint.  

Previously, we introduced the \hlsfml~library to facilitate the deployment of DNN models on L1T systems~\cite{hls4ml}. The aim of that work was to establish an automatic workflow to convert a given DNN model into an electronic circuit, evaluated on an FPGA through a fully-on-chip firmware implementation. The workflow consists of converting a given NN model into an expertly written C++ code, which is then converted to an FPGA firmware by a High Level Synthesis (HLS) tool (e.g., Xilinx Vivado HLS). In Ref.~\cite{hls4ml}, we demonstrated how a  DNN model for jet identification at the LHC could be compressed and quantized, to run on an FPGA with $\SI{75}{ns}$ latency.

In this work, we present an extension of the \hlsfml~library to also support BDTs.
As shall be seen in the following Sections~\ref{sec:hls4ml} and~\ref{sec:results}, the BDT implementation in FPGAs is capable of achieving similar performance to a DNN, with a relatively lightweight usage of device resources.
The critical FPGA resource for BDTs is Look Up Tables (LUTs), whereas the availability of DSPs for multiplication is the limiting factor for DNNs. 
Given this, the BDT can be seen as a lightweight solution which is complementary to a DNN.

Another motivation for the introduction of BDTs is the need to support the legacy of the LHC Run II: as of today, BDTs are still the most commonly used ML algorithm for LHC experiments. For instance, the LHCb Collaboration makes extensive use of BDTs (as well as neural networks) in their trigger, which runs in software only. To accelerate the computation, a binned BDT method, Bonsai BDT, is used \cite{bonsai_bdt}. 

BDTs remain a particularly appealing solution for use in the earliest processing stages at LHC experiments, thanks to their good performance with relatively low computational cost. The first use case of an ML technique in the L1T of an LHC experiment was a BDT used to perform a regression of muon $p_\text{T}$ for the CMS L1T endcap muon trigger~\cite{cms_l1t_bdt}.
The technique gave a three-times reduction in rate for the trigger threshold compared to the previous approach, removing unwanted low $p_\text{T}$ muons. An external DRAM of $\SI{1.2}{GB}$ was used as a look-up-table (LUT) to store the pre-computed BDT output for every variation in the input variables. The LUT was filled offline and queried with low latency online. The solution proposed in this paper would allow an on-chip implementation going beyond a full-LUT approach. 

%In this work, we pursue the same paradigm of offline training with online inference, but implement the BDT operations in the FPGA logic.
%In particular, the implementation targets extremely low inference latencies, compatible with use in the L1T of LHC experiments.

Other works have implemented ensembles of Decision Trees (BDTs and Random Forests) for FPGAs \cite{fpga_bdt_mem, fpga_bdt_distributed, fpga_bdt_1, fpga_bdt_2, fpga_bdt_3}. These generally target applications of FPGA accelerated inference in a combined CPU-FPGA system, where the relevant performance goals are throughput and energy consumption. Further, the use of external memories and traversal over trees by fetching nodes from memory gives these approaches flexibility and scalability.
The work of \cite{fpga_bdt_mem} and \cite{fpga_bdt_distributed}, in particular, is designed to be scalable to very large ensembles in a way that the implementation in this paper is not.
In the context of targeting LHC triggers, however, the main performance goal is of extremely low latency, and secondly to maintain a modest resource usage.

%This paper is organized as follows: Section~\ref{sec:hls4ml} describes the building of BDT models in \hlsfml. Implementation and performances are discussed in Section~\ref{sec:results}. Conclusions are given in Section~\ref{sec:outlook}.

% %%%%%%%%%%%%%%%%%%%%%%%%%%%%%%%%%%%%%%%%%%%%%%%%%%%%%%%%%%%%%%%%%%%%%%%%%%%%%%%%%%%%%%%%%%%%%%%%
% % H L S 4 M L
% %%%%%%%%%%%%%%%%%%%%%%%%%%%%%%%%%%%%%%%%%%%%%%%%%%%%%%%%%%%%%%%%%%%%%%%%%%%%%%%%%%%%%%%%%%%%%%%%
%\clearpage
\section{Building Boosted Decision Trees with \hlsfml}
\label{sec:hls4ml}
\begin{figure}[tb!]
    \centering
    \includegraphics{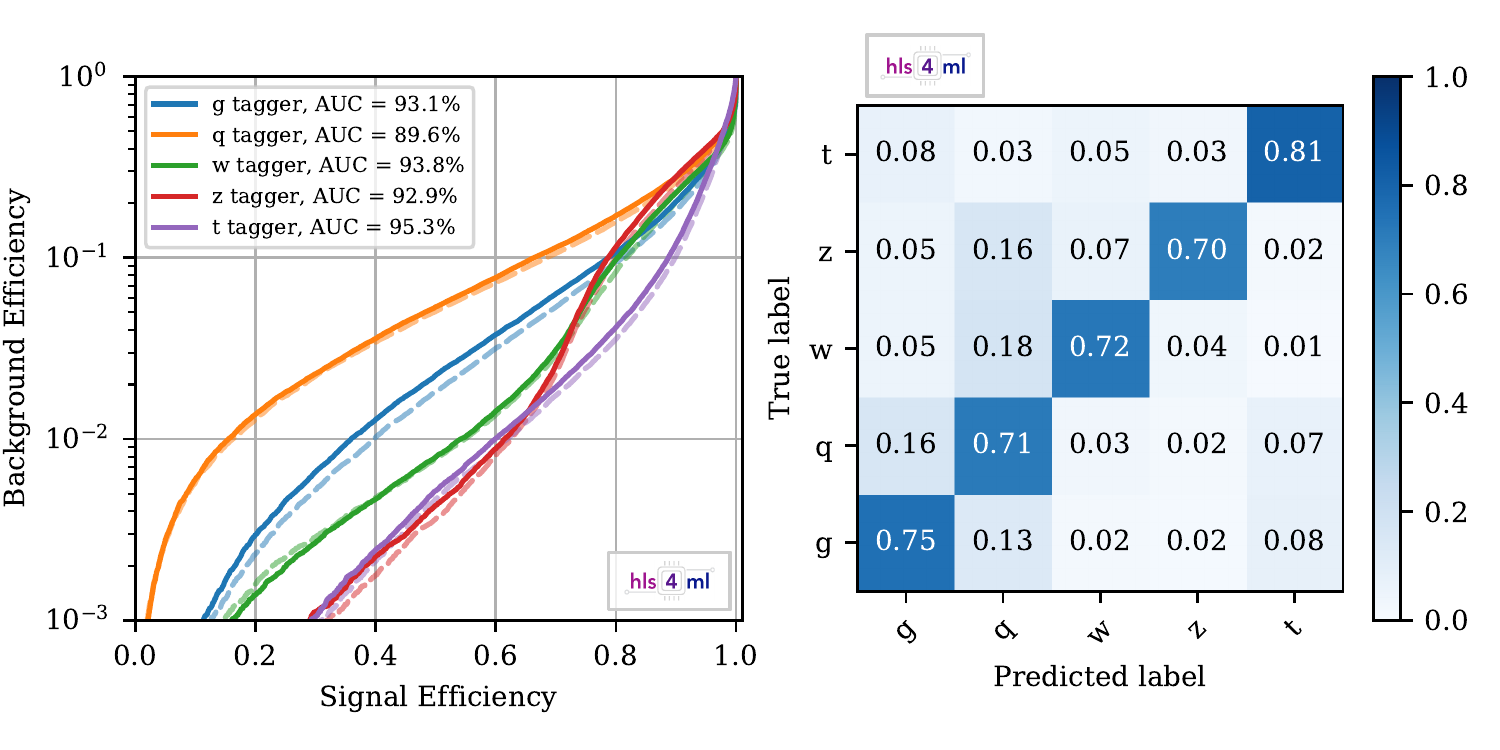}
    \caption{Left: The solid curves show signal efficiency vs. misidentification rate using a BDT with 100 trees of depth 4 for the five jet classes: gluon, quark, $W$ boson, $Z$ boson, and top quark. The dashed curves show the performance of the 3 layer MLP from~\cite{hls4ml}. Right: confusion matrix for the BDT.}
    \label{fig:roc}
\end{figure}

In the previous work on translation of neural networks to FPGA firmware with \hlsfml, we presented a demonstration data set for discrimination of quarks ($q$), gluons ($g$), $W$ and $Z$ bosons, and top ($t$) jets~\cite{pierini_maurizio_2020_3602260}. The data consist of a set of 16 physics-motivated high-level features, representing information of the event jet substructure. With this information at hand, one can distinguish traditional single-prong $q$ and $g$ jets from two- ($W$ and $Z$) and three-prong jets

This problem is typical of searches for physics beyond the standard model at ATLAS and CMS. To our knowledge there is no algorithm currently employed in the L1T systems of these two experiments that exploits this kind of {\it substructure} information to select events with multi-prong jets. This data set provides a benchmark on which to evaluate the classifier performance and its realisation in FPGA implementation as an example application for the L1T.
We use the same data set in this work to prepare a classifier, this time a BDT. 

We performed the BDT training using the \texttt{scikit-learn} package \cite{scikit-learn}, randomly splitting the data set into training (80\%) and testing (20\%) partitions.
A BDT with 100 trees and a maximum depth of 4 was found to give similar performance to the DNN model trained on the same data set, providing a useful point of comparison.
The cross-entropy loss function was used.

The resulting receiver operating characteristic (ROC) curve is shown in Figure~\ref{fig:roc}, displaying the background misidentification efficiency (false-positive rate) as a function of the signal efficiency (true-positive rate) for five jet selectors, defined using the five scores returned by the BDT for the five jet categories. Overall, the trained BDT reaches state-of-the-art discrimination performance, with a small performance loss with respect to the DNN model of Ref.~\cite{hls4ml}.

%{\bf should we add some characterization of the model in terms of number of operations that it requires? Maybe comparing it to the DNN to show the trade-off between performance and resources? What about:}
The operations used in inference of a BDT are very different from those used for a neural network.
While a (fully connected) neural network comprises a series of matrix-vector products and evaluations of non-linear activation functions, the BDT inference involves evaluating decision paths over many decision trees.
This tree traversal requires comparisons against thresholds, effectively partitioning the feature space.
In terms of the number of parameters, the trained BDT with 100 trees of depth 4 is summarised by 7,500 threshold values, and 8,000 scores.
The fully connected neural network presented in~\cite{hls4ml}, with the same 16 inputs, 5 outputs, and three hidden layers of 64, 32, and 32 neurons, has 4,389 trainable parameters. 
A BDT is only able to make cuts orthogonal to the feature axes, while the activation functions of a neural network add non-linearity to the classification.

We use this model as a benchmark example to show the use of \hlsfml~to derive an FPGA firmware implementation.

% %%%%%%%%%%%%%%%%%%%%%%%%%%%%%%%%%%%%%%%%%%%%%%%%%%%%%%%%%%%%%%%%%%%%%%%%%%%%%%%%%%%%%%%%%%%%%%%%
% % R E S U L T S
% %%%%%%%%%%%%%%%%%%%%%%%%%%%%%%%%%%%%%%%%%%%%%%%%%%%%%%%%%%%%%%%%%%%%%%%%%%%%%%%%%%%%%%%%%%%%%%%%
%\clearpage
\section{Implementation and Performance}
\label{sec:results}
\begin{figure}[t!]
    \centering
    \includegraphics[width=0.8\textwidth]{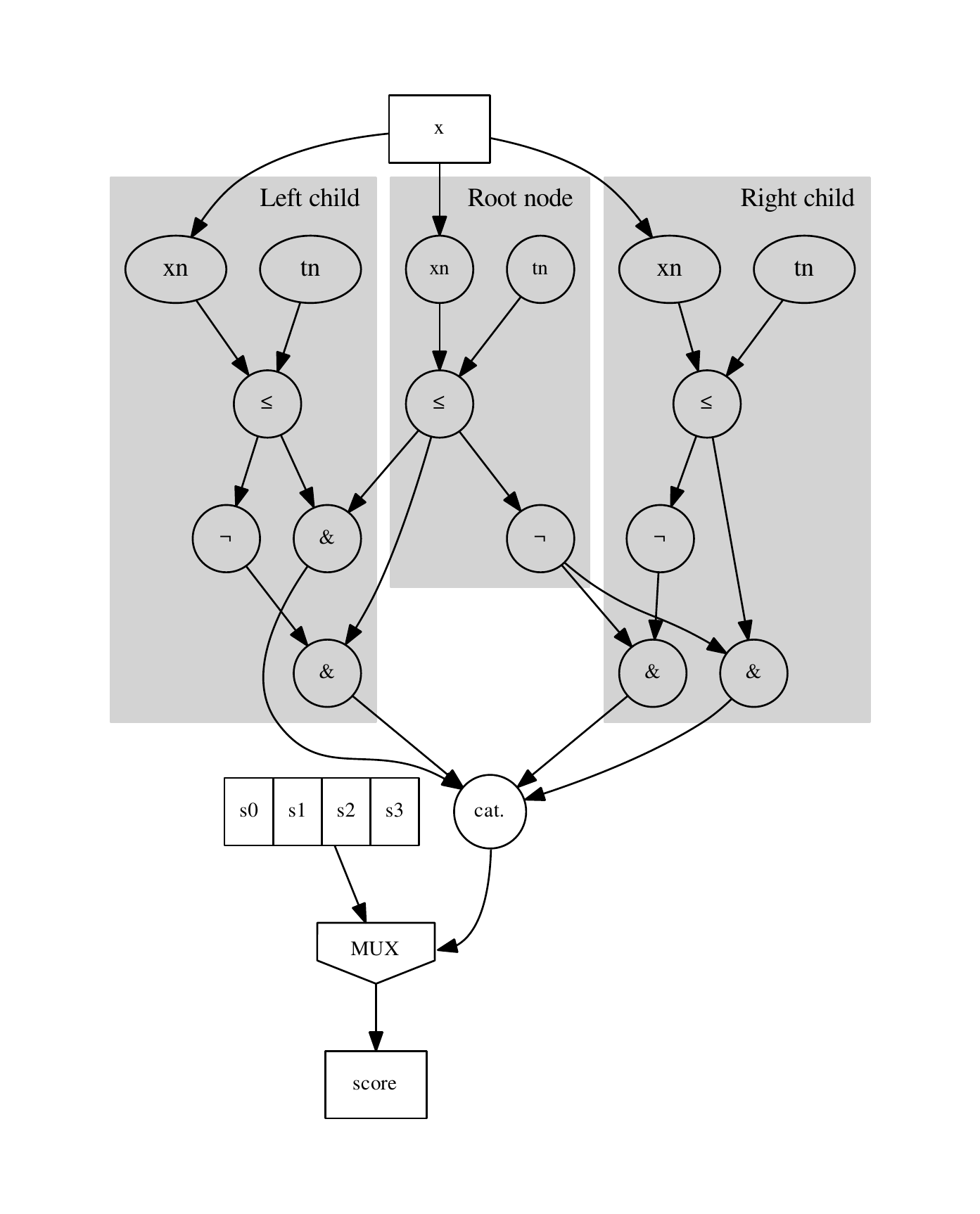}
    \caption{Schematic of the implementation of decision trees in \hlsfml, showing a single tree with depth of 2. The $x$ are the node features, and $t$ the thresholds. The `$\neg$' is the unary `not' operator, and `\&' the binary `and'. The Boolean leaf activations are concatenated and used to address a look-up-table of output scores.\label{fig:dt_schematic}}
\end{figure}

\subsection{FPGA implementation}

Decision Trees in \hlsfml~are implemented as an unrolled tree of decisions, as illustrated in Figure~\ref{fig:dt_schematic}.
Each node in the tree performs a comparison of one of the input features against a constant threshold, learned in training.
These thresholds are statically fixed in the logic of the FPGA firmware, rather than being fetched from an external memory.
Nodes pass the results of the comparison (true or false) to their children. %-- inverted for one child.
The decision path is then encoded by the series of Boolean values propagated along the nodes.
By construction, only a single leaf node can be activated, and the index of the active leaf is use to address a small look-up-table containing the tree scores for each path.
These scores map to the probability that the given input features correspond to a certain class.

The score of the BDT ensemble is the sum of scores of all of the decision trees. Since each decision tree is independent, a high degree of parallelisation is possible in the FPGA. The sum is performed with a balanced adder tree. The implementation of BDTs in \hlsfml~targets low latency applications, such as LHC hardware triggers, by executing all trees, and all decisions within each tree, in parallel.

We developed two code implementations, both targeting the architecture described. The first uses Xilinx's Vivado HLS, written in C++, and the second is developed at the Register-Transfer Level (RTL), using VHDL. Generally, an RTL implementation does not benefit from some of the features of Vivado HLS, such as automatic pipelining depending on the target clock frequency, and easy loop rolling/un-rolling.
However, the RTL implementation synthesises to more reliable results for `large' BDTs, as will be seen in the Section~\ref{sec:performance}. Both implementations are fully pipelined, capable of an `initiation interval' of 1 clock cycle.

A trained BDT, with specific features, thresholds and scores for each tree, can be evaluated with the FPGA implementation described above using \hlsfml.
Models trained and exported from the \texttt{scikit-learn}, \texttt{xgboost}~\cite{xgboost}, and \texttt{TMVA}~\cite{tmva} packages are supported.
From the FPGA code produced, which is either using Vivado HLS or VHDL, the user is then able to run the usual FPGA vendor workflow to integrate the BDT into a specific project and compile to a bitfile.

\subsection{Varying the precision}
\label{sec:precision}
The generic, programmable-logic cells in FPGAs support completely customised data representations.
Floating point types are supported, but generally require more resources, latency, and achieve lower clock frequencies than integer types.
The fixed point representation uses integer operations, but with a radix point in the number to represent fractional values.
In the FPGA, any bitwidth and radix position may be used.
A narrower bitwidth will enable smaller resource usage.

\begin{figure}[t!]
    \centering
    \includegraphics{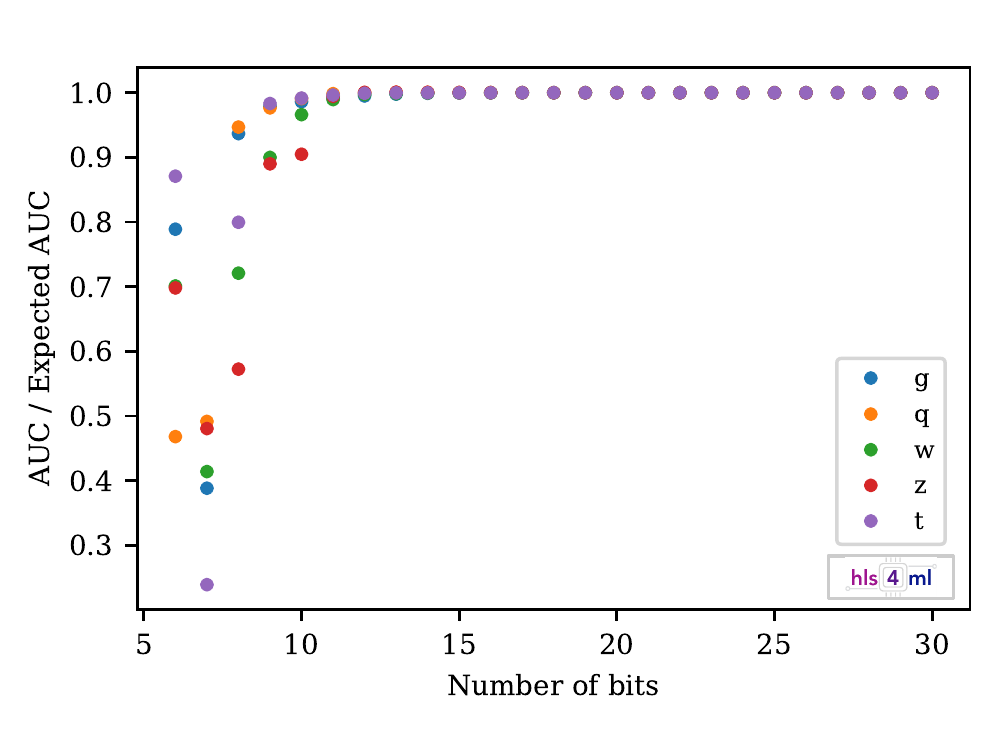}
    \caption{The ratio of Area Under the Curve (AUC) obtained from the fixed point implementation to the AUC expected from the floating point software, as a function of the fixed point bitwidth used, for each of the five tag categories. The ratio saturates at around 15 bits.
    \label{fig:w_vs_auc}}
\end{figure}

The trade-off for using narrower bitwidths is a loss of precision.
%For a classifier, this can lead to misclassification ...
The loss of discriminating power is investigated by measuring the ratio of the AUC obtained testing with fixed point representation to the area under the ROC curve (AUC) from the original floating point, and shown in Figure \ref{fig:w_vs_auc} as a function of the bitwidth, for the benchmark jet-classification BDT introduced in Section~\ref{sec:hls4ml}.
The number of integer bits was kept at 4 for all bitwidths, as required by the range of the features and scores in the data to avoid overflow.
A significant reduction in AUC is seen for the smallest width of 6.
The AUC with fixed point variables reaches 99\% of the AUC with floating point for all taggers with 11 bits. The consequences in terms of resource savings are discussed in the next section.

\subsection{Performance and cost}
\label{sec:performance}

\begin{table}[b!]
\centering
\begin{tabular}{|l|r|r|r|r|}
\hline

     Resource &  LUTs & FFs & DSPs & BRAMs\\
     \hline
     \hline
     Number Used & 96148 & 42802 & 0 & 0 \\
     \hline
     Percentage of VU9P & 8.1 & 1.8 & 0 & 0 \\
     \hline
\end{tabular}
\caption{Resource usage of the BDT with 100 trees of depth 4.
\label{tab:bdt_200_4_util}}
\end{table}

We studied the FPGA resource utilisation and inference latency using BDTs trained on the jet classification task described in Section~\ref{sec:hls4ml}.
These metrics are expected to vary with the number of trees and their depth. 
Other hyperparameters, while having an impact on the classification performance, do not affect these FPGA performance metrics.
All HLS evaluations of BDTs were built for a Xilinx \texttt{vu9p-flgb2104-2L-e} FPGA at $\SI{200}{MHz}$ target clock frequency.
FPGAs of this size or similar could be used in future LHC upgrades, and would generally be used to execute several algorithms (including feature pre-processing) as well as any ML inference.
All features, thresholds, and scores were encoded with 18 bits, which is sufficient to achieve identical classification results to the \texttt{scikit-learn} original, as was shown in Section~\ref{sec:precision}.

The resource utilisation of LUTs, FFs, DSPs, and BRAMs for the benchmark BDT with 100 trees and a depth of 4 is shown in Table~\ref{tab:bdt_200_4_util}. 
This utilisation is reported after running the logic synthesis step with the VHDL implementation.
The inference latency for this ensemble is 12 clock cycles, corresponding to $\SI{60}{ns}$ execution time at the chosen target clock frequency.
This is compatible with the requirements for use in the L1T system.

% Resources
\begin{figure}[t!]
\centering
\includegraphics[width=\textwidth]{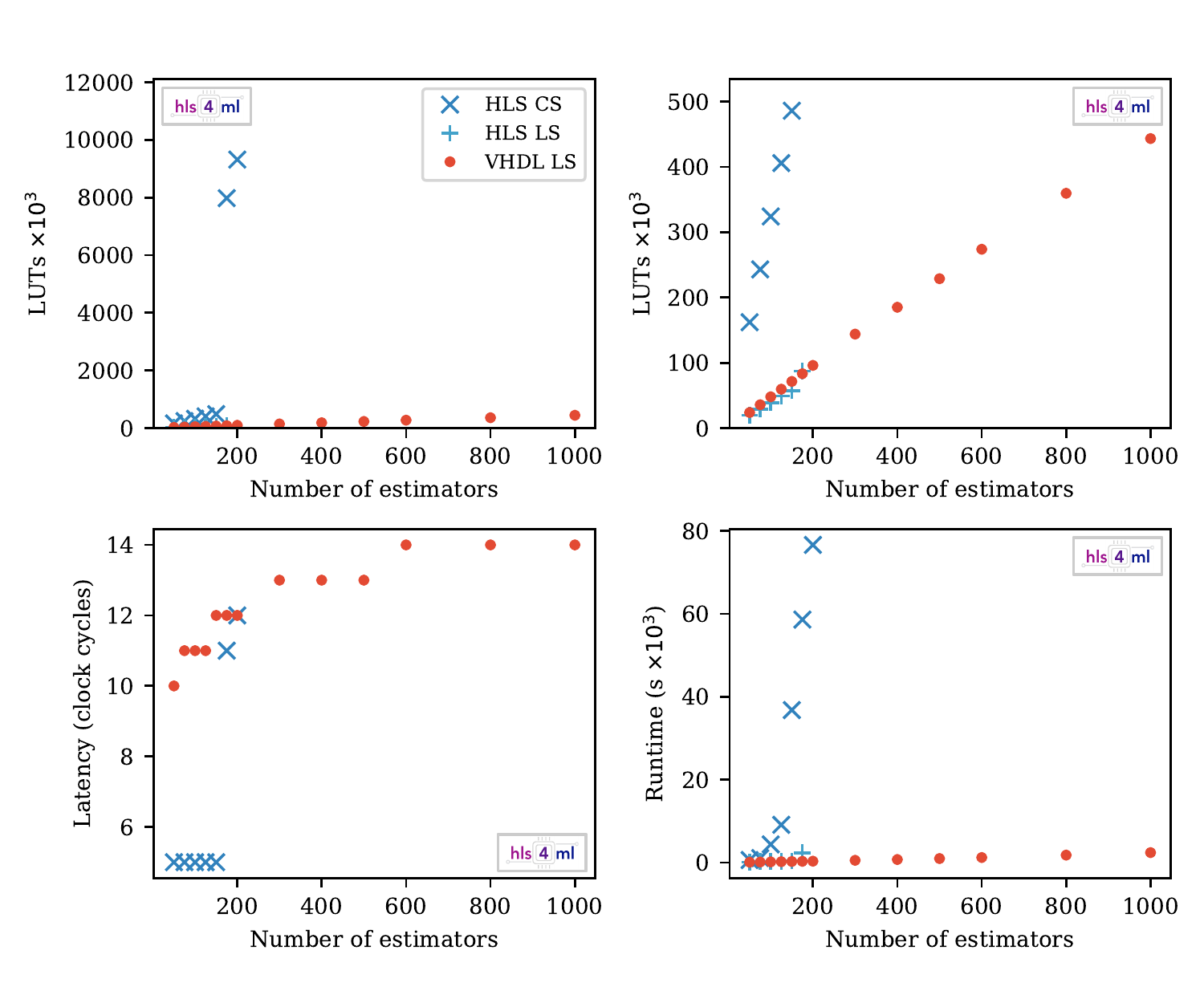}
\caption{Dependence of LUT usage (top row), inference latency (bottom left), and synthesis time (bottom right) on the number of estimators (trees) of the BDT, with depth fixed at 3. The top right plot is a view of the same data as the top left, with reduced range. Different stages of synthesis are shown: C-Synthesis estimate of HLS (HLS CS), utilisation report after Logic Synthesis step of RTL produced by HLS (HLS LS), and utilisation report after Logic Synthesis of the VHDL implementation (VHDL LS).}
\label{fig:n_estimators_scan}
\end{figure}

Figure~\ref{fig:n_estimators_scan} shows the variation in resource usage with the number of estimators, $n_\text{e}$, of the BDT, with the depth fixed at 3. 
For multiclass classification, each estimator uses as many trees as the number of classes, in this case of the jet classification dataset, five.
Only one tree per estimator would be used for a binary classification problem, reducing the resource cost by a factor five.
For the VHDL implementation, the utilisation is reported after logic synthesis with Vivado.
For the HLS implementation, the Vivado HLS resource estimate after C-synthesis is reported, as well as the result after executing logic synthesis on the produced RTL with Vivado.
The HLS estimate of LUT and FF usage tend to be larger than the eventual usage after the full synthesis and implementation workflow.

% Resources
\begin{figure}[t!]
\centering
\includegraphics[width=\textwidth]{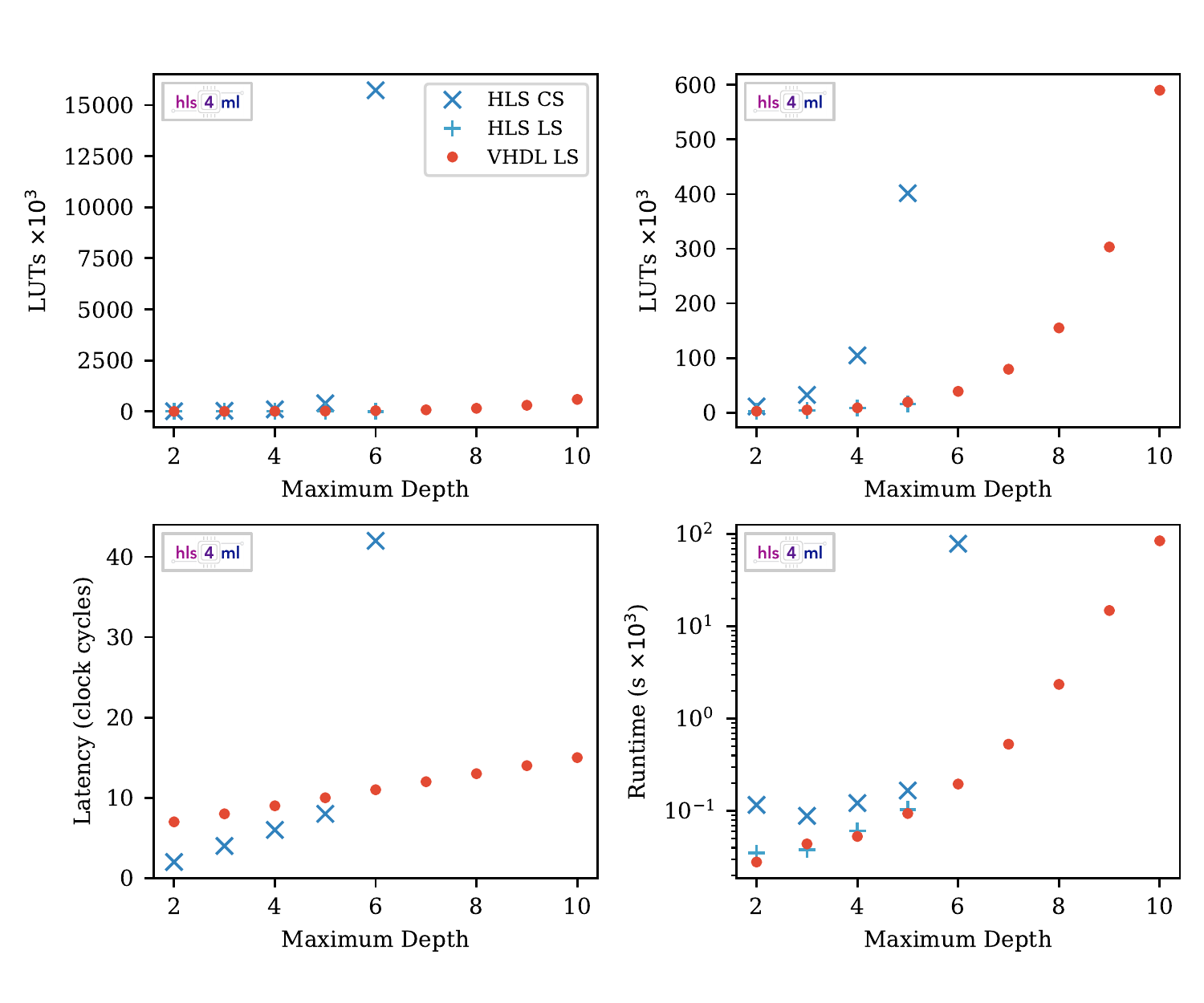}
\caption{Dependence of LUT usage (top row), inference latency (bottom left), and synthesis time (bottom right) on the maximum depth of the BDT, with 10 estimators. The top right plot is a view of the same data as the top left, with reduced range. Different stages of synthesis are shown: C-Synthesis estimate of HLS (HLS CS), utilisation report after Logic Synthesis step of RTL produced by HLS (HLS LS), and utilisation report after Logic Synthesis of the VHDL implementation (VHDL LS).\label{fig:max_depth_scan}}
\end{figure}

Up to $n_\text{e} = 150$, the LUT utilisation for both implementations increases linearly, with the HLS implementation using slightly fewer than the VHDL version (referring to the utilisation reported after Vivado synthesis).
With $n_\text{e} > 150$, the LUT usage of the HLS implementation increases dramatically, and the Vivado synthesis of the produced RTL also yields poor results.
In this regime, the LUT usage of the VHDL implementation continues to increase linearly with $n_\text{e}$.

The inference latency of the VHDL implementation increases logarithmically with $n_\text{e}$, as the depth of the balanced add-tree used to sum tree score increases.
The HLS implementation inference latency is more constant, as HLS packs the add-tree into a single cycle for most ensemble sizes.
For $n_\text{e} > 150$ the latency of the HLS result increases significantly.
The VHDL implementation latency is typically longer than the latency achieved by the HLS.
The VHDL is pipelined to achieve timing closure at higher clock frequencies than the $\SI{200}{MHz}$ target used for the HLS.

The time taken to synthesise the BDT increases linearly with $n_\text{e}$ for the VHDL implementation, taking 40 minutes for the 1000 estimators ensemble.
The HLS C synthesis time increase exponentially with the number of estimators, with synthesis for 200 estimators taking 21 hours.
Vivado synthesis times for the HLS RTL output are significantly faster than the HLS C Synthesis which must run before, and increase linearly with the number of estimators.

Figure ~\ref{fig:max_depth_scan} shows the dependence of the same FPGA performance metrics on the maximum depth of the BDT, with $n_\text{e}$ fixed at 10.
The LUT usage increases exponentially with depth, with each additional layer in the trees adding as many nodes as there are above it.
As before, the HLS estimate of the LUTs is high compared with the report after synthesising the produced RTL with Vivado.
The LUT usage of the VHDL and Vivado-synthesized HLS are very similar, until at maximum depth of 6, the HLS implementation resource usage suddenly increases.
At the same point, the latency and synthesis time drastically increase.
The latency of the VHDL implementation increases linearly, with one extra clock cycle per depth.
Synthesis time increases exponentially with depth, with the synthesis for a depth of 10 taking 27 hours.

\subsection{Resource model}
\begin{figure}[t]
    \centering
    \includegraphics{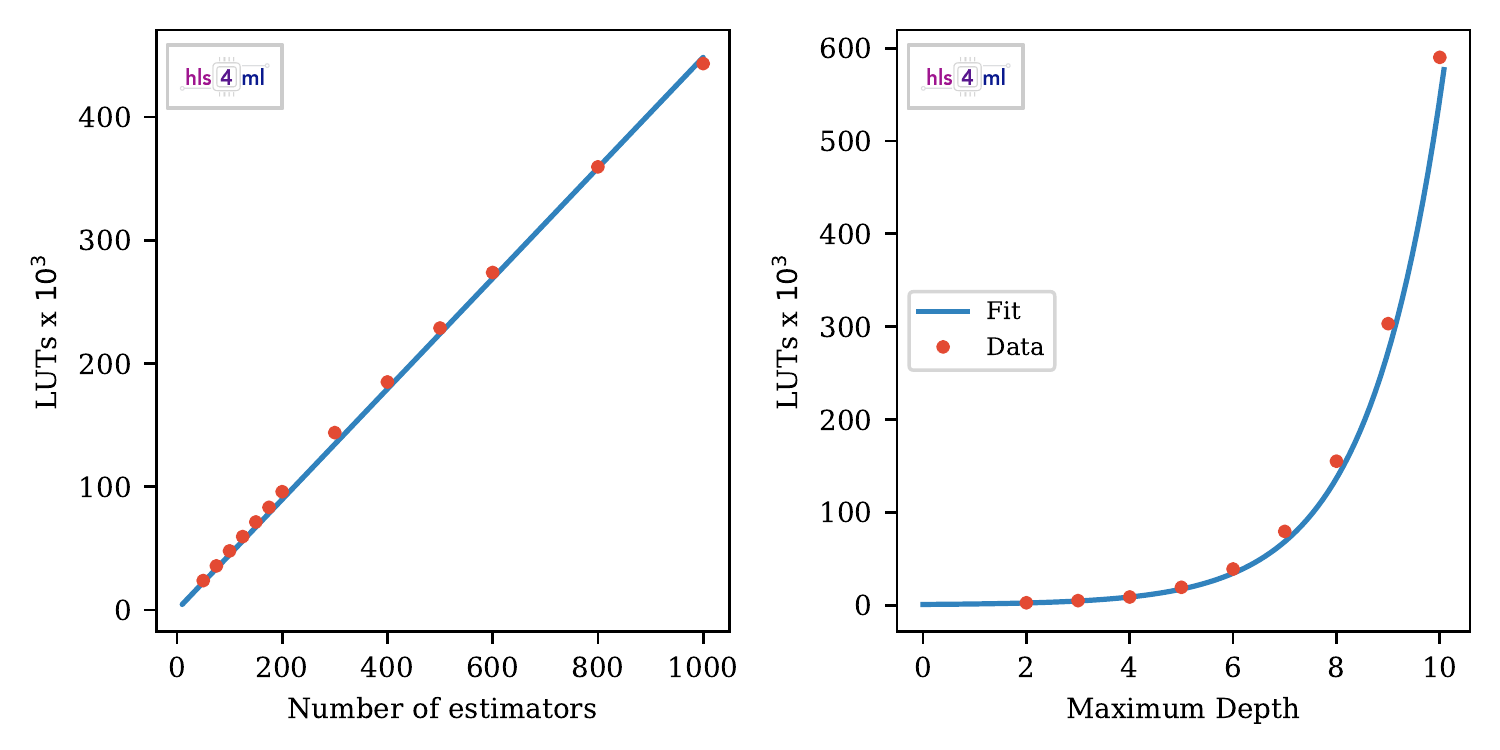}
    \caption{Comparison of LUT usage between measured results (points) and resource usage model (curve). Left: as a function of number of estimators with depth fixed at 4. Right: as a function of the depth, with number of estimators fixed at 10.}
    \label{fig:scaling}
\end{figure}

Given the expected scaling of LUTs with model hyperparameters -- linear with $n_\text{e}$ and exponentially (base two) with depth -- we analytically describe the resource usage using the following relation:
\begin{equation*}
    r = k_0 \cdot n_\text{e} + k_1\cdot n_\text{e} \cdot 2^d, 
\end{equation*}
where $r$ is the resource usage (LUTs), $n_\text{e}$ the number of estimators, $d$ the tree depth, and $k_0$, $k_1$ are unknown constants.
The term linear in $n_\text{e}$ represents the resource of the adder-tree which grows with the number of trees.
The term linear in $n_\text{e}$ and exponential with $d$ represents the logic used for the trees, of which there are $n_\text{e}$, while the number of decision nodes doubles at each layer in depth.
Other hyperparameters -- such as the loss function, learning rate, and number of features -- may impact the classification performance of the model, but would not affect the resource usage.
A fit to the measurements of trained and synthesised BDTs using the VHDL implementation was performed, yielding:
\begin{equation*}
    r = 22 \cdot n_\text{e} + 53 \cdot n_\text{e} \cdot 2^d.
\end{equation*}
All features, thresholds and scores were encoded with 18 bits.
Figure \ref{fig:scaling} shows this scaling model over the measured BDT results used for the single-parameter scans in Figures~\ref{fig:n_estimators_scan} and ~\ref{fig:max_depth_scan}, showing good agreement.

\begin{figure}[t]
\centering
\includegraphics{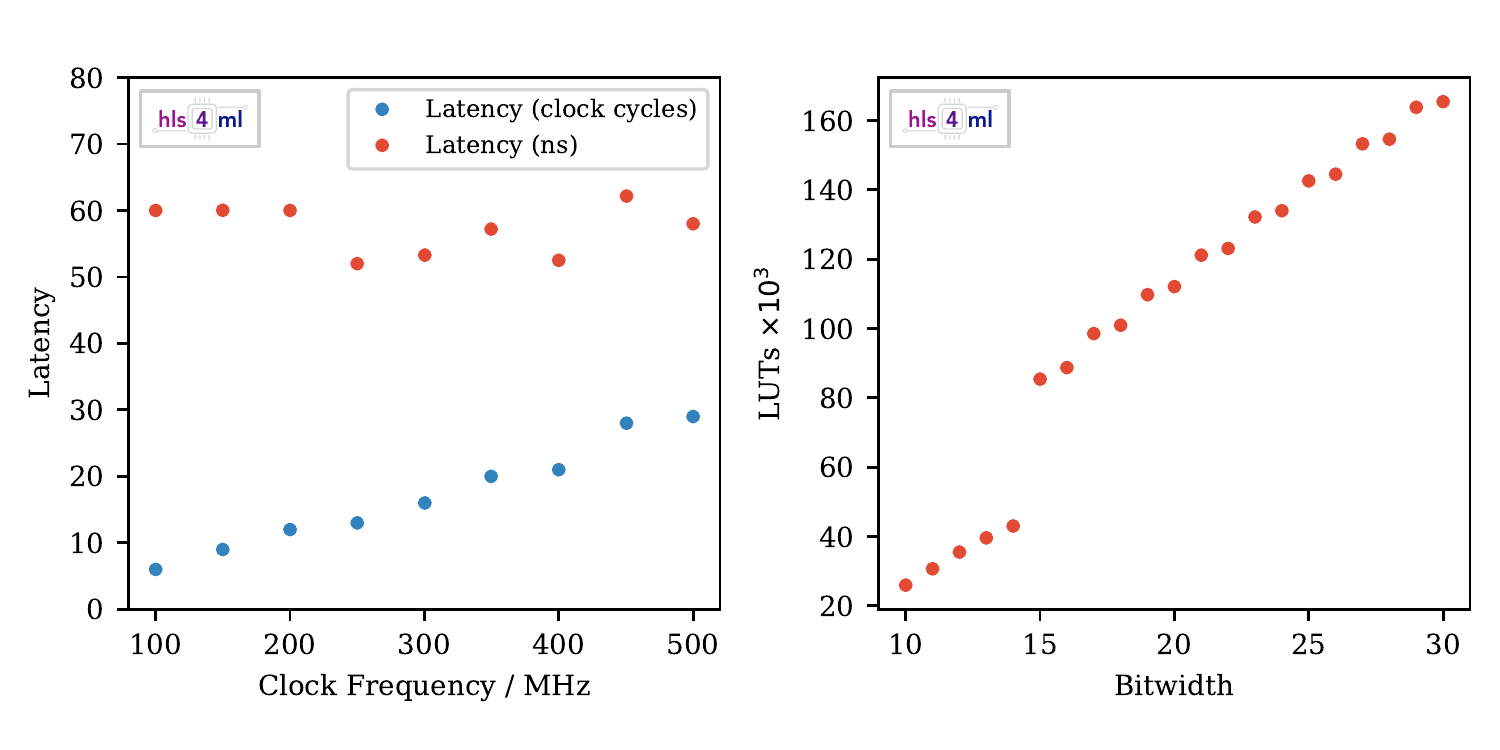}
\caption{Left: Latency -- in clock cycles and nanoseconds -- of the benchmark BDT model with 100 trees of depth 4, as a function of the target clock frequency. Right: Resource usage as a function of the bitwidth used for all features, thresholds and scores.\label{fig:frequency_v_latency_and_precision_scan}}
\end{figure}

\subsection{Varying clock frequency and precision}
% Latency scan
Vivado HLS automatically pipelines FPGA designs, according to the target clock period specified by the developer.
When using the HLS workflow, the \hlsfml~library allows the user to choose a target clock period for the BDT model.
%Since the design is pipelined, a higher clock frequency allows for a greater inference throughput.
Generally, a faster target clock frequency requires more pipeline stages, so more clock cycles will be needed to perform the inference.
The left plot of Figure~\ref{fig:frequency_v_latency_and_precision_scan} shows the pipeline depth increasing with target clock frequency from 6 clock cycles at $\SI{100}{MHz}$ to 29 cycles at $\SI{500}{MHz}$.
The single inference latency in nanoseconds (the product of the latency in clock cycles and the clock period) is relatively constant with the target clock frequency.
The lowest single inference latency  is $\SI{52}{ns}$ at $\SI{250}{MHz}$ while the highest is $\SI{62.2}{ns}$ at $\SI{450}{MHz}$.
Using a higher clock frequency will achieve overall faster inference when classifying several input feature vectors, since the initiation interval is 1 in all cases.

The variation of resource usage with the bitwidth is shown in the right plot of Figure~\ref{fig:frequency_v_latency_and_precision_scan} for LUTs, the dominant resource used for BDTs.
Four integer bits were used in all cases, as in Figure~\ref{fig:w_vs_auc}.
The increase in resource usage with bitwidth is approximately linear, but with a significant step change transitioning from 14 to 15 bits.

%%%%%%%%%%%%%%%%%%%%%%%%%%%%%%%%%%%%%%%%%%%%%%%%%%%%%%%%%%%%%%%%%%%%%%%%%%%%%%%%%%%%%%%%%%%%%%%%
% S U M M A R Y
%%%%%%%%%%%%%%%%%%%%%%%%%%%%%%%%%%%%%%%%%%%%%%%%%%%%%%%%%%%%%%%%%%%%%%%%%%%%%%%%%%%%%%%%%%%%%%%%
%\clearpage
\section{Summary and Outlook}
\label{sec:outlook}
We presented the implementation of BDT conversion to FPGA firmware in the \hlsfml~library. Taking as an example a multiclass classification problem from high energy physics (the identification of boosted jets based on substructure information), we show how a state-of-the-art algorithm could be deployed on an FPGA with a typical inference time of 12 clock cycles (i.e., $\SI{60}{ns}$ at a clock frequency of $\SI{200}{MHz}$).
%We discuss the optimization options provided by the library, and how they could be used to reduce the resource consumption. 
We discussed the dependence of the FPGA resource usage and inference latency upon the model hyperparemeters, presenting a model which predicts the resource usage well.
We compared an HLS-based implementation to a VHDL one, as a function of the model size. Both the workflows are supported in \hlsfml.
The presented workflow provides a resource effective alternative to Neural Network deployment, which we discussed in a previous publication~\cite{hls4ml}.
Compared to a Neural Network applied to the same problem, a BDT is able to achieve very similar performance, with a comparable inference latency.
The implementation of BDTs in the FPGA utilises LUTs most heavily, while the Neural Network predominantly uses DSPs. 
This functionality of the  \hlsfml~library could support an efficient deployment of algorithms analogous to that described in Ref.~\cite{cms_l1t_bdt}, which took data at the CMS experiment during the LHC Run II.

%%%%%%%%%%%%%%%%%%%%%%%%%%%%%%%%%%%%%%%%%%%%%%%%%%%%%%%%%%%%%%%%%%%%%%%%%%%%%%%%%%%%%%%%%%%%%%%%
% Acknowledgements
%%%%%%%%%%%%%%%%%%%%%%%%%%%%%%%%%%%%%%%%%%%%%%%%%%%%%%%%%%%%%%%%%%%%%%%%%%%%%%%%%%%%%%%%%%%%%%%%
\section*{Acknowledgements}
\label{sec:acknowledgements}
We acknowledge the Fast Machine Learning collective as an open community of multi-domain experts and collaborators. This community was important for the development of this project. 
\initials{M.P.}, \initials{S.S.}, \initials{V.L.} and \initials{J.N.} are supported by the European Research Council (ERC) under the European Union's Horizon 2020 research and innovation program (grant agreement n$^o$ 772369).
\initials{S.J.}, \initials{R.R.}, and \initials{N.T.} are supported by Fermi Research Alliance, LLC under Contract No. DE-AC02-07CH11359 with the U.S. Department of Energy, Office of Science, Office of High Energy Physics.
P.H. is supported by a Massachusetts Institute of Technology University grant. 
\initials{Z.W.} is supported by the National Science Foundation under Grants No. 1606321 and 115164.

\bibliographystyle{mine}
\bibliography{bibliography}

\end{document}